\documentclass[11pt,twoside]{article}

\usepackage{asp2006}
\usepackage{graphicx}

\begin{document}
\title{Gaia and the Astrometry of Giant Planets}   
\author{M. G. Lattanzi and A. Sozzetti}   
\affil{INAF - Ossservatorio Astronomico di Torino, I-10025 Pino Torinese, Italy}    

\begin{abstract} 
Scope of this contribution is twofold. First, it describes the
potential of the global astrometry mission Gaia for detecting and
measuring planetary systems based on detailed double-blind mode
simulations and on the most recent predictions of the satellite's
astrometric payload performances(launch is foreseen for late
Summer 2012). Then, the identified capabilities are put in context
by highlighting the contribution that the Gaia exoplanet
discoveries will be able to bring to the science of extrasolar
planets of the next decade.
\end{abstract}


\section{Gaia in a nutshell}
The Gaia mission is the new global, all-sky, astrometric
initiative of the European Space Agency with a launch possibility
occurring in late Summer 2012. A Soyuz-Fregat launcher will take
the Gaia module to a transfer orbit, which in one month will allow
the satellite to reach its operational environment on a Lissajous
orbit at Sun-Earth L2, 1.5 million kilometers away from Earth.
During its 5 years of operational lifetime (with the possible
extension of an extra year) Gaia will monitor all point sources in
the visual magnitude range $6-20$ mag, a huge database of some
$\sim10^9$ stars, a few million galaxies, half a million quasars,
and a few hundred thousand asteroids.\\
As for the observing strategy, Gaia's mode of operation has
adopted the principles successfully experimented with the
Hipparcos mission (ESA 1997). In particular, it will continuously
scan the sky implying that all detected objects, irrespective of
their magnitudes, are observed for the same amount of time during
each field-of-view crossing, with mission-end observing time
mainly depending on ecliptic latitude (\cite{lindegren2010}).\\
In this way, it is anticipated that Gaia will determine the five
basic astrometric parameters (two positional coordinates, two
proper motion components, and the parallax) for all objects, with
end-of-mission (sky-averaged) precision between 7-25 $\mu$as
(microarcsec) down to the Gaia magnitude $G$\footnote{close to
Johnson's $R$.} $=15$ mag and a few hundred $\mu$as at $G=20$ mag,
depending on color. Objects redder than $V-I= 0.75$ are expected
to have better astrometry, while that of extreme blue targets is
estimated to degrade by a factor of two.

A combination of an ambitious science case, wishing to address
breakthrough problems in Milky Way astronomy, and lessons learned
from the Hipparcos experience brought European astronomers to
realize that modern astrometry cannot do without
spectrophotometry. That is why Gaia's astrometry is complemented
by on-board spectrophotometry and (only for objects brighter than
$G=17$) radial velocity information. These data have the precision
necessary to quantify the early formation, and subsequent
dynamical, chemical and star formation evolution of our Galaxy.
The broad range of crucial issues in astrophysics that will be
addressed by the wealth of the Gaia data is summarized by e.g.,
Perryman et al. (2001). One of the relevant areas on which the
Gaia observations will have great impact is the astrophysics of
planetary systems (e.g., Casertano et al. 2008), in particular
when seen as a complement to other techniques for planet detection
and characterization (e.g., Sozzetti 2009).

\section{Project organization}

The main partners behind the Gaia project are: i) the European
Space Agency (ESA), which has the overall project responsibility
for funding and procurement of the satellite, launch, and
operations. Of interest is the fact that in this case satellite
procurement includes the payload and its scientific instruments,
unlike ESA's other science missions for which scientific
instruments are usually PI-lead and funded (or, at least,
co--funded) by participating national space agencies; ii)EADS
Astrium, who was selected in 2006 as the prime industrial
contractor for designing and building the satellite according to
the scientific and technical requirements formulated to fulfill
the mission science case as approved at time of selection (ESA
2000); the Gaia Data Processing and Analysis Consortium (DPAC),
charged with designing, implementing and running a complete
software system for the scientific processing of the satellite
data, resulting in the 'Gaia Catalogue' a few years after the end
of the operational (observation) phase.

DPAC was formed in 2006 in response to an 'Announcement of
Opportunity' issued by ESA. The Consortium currently lists nearly
400 individual members in more than 20 countries, including a team
at the European Space Astronomy Center near Madrid (Spain). Six
data processing centers participate in the activities of the
consortium, which is organized in eight 'coordination units' each
responsible for the development of one part of the software like,
e.g., {\it simulations}, {\it core processing} (global
astrometry), {\it photometry}, and {\it non-single stars}, the
unit devoted to the processing of astrometrically "noisy" stars,
which will include potential planetary systems. Most of the 
financial support is provided by ESA (for the team at ESAC) and by the
various national space agencies through a legally binding long--term 
funding agreement, a real first for ESA run missions.

There will be no proprietary periods for the scientific
exploitation of the data. The final Gaia catalogue will be
produced and immediately delivered to the astronomical community
worldwide as soon as ESA and DPAC will agree on the processed data
having reached the targeted (science) quality. This catalogue
is expected to be ready three years after the end of operations.
Finally, intermediate releases, of some provisional results, are
planned after a few years of observations.

More information can be found in Lindegren (2010), while other
organizational details and the latest news on payload and
satellite developments are available on the Gaia web pages at
http://www.rssd.esa.int/gaia/.

\section{Gaia and extrasolar planets}

\subsection{What will Gaia see?}

As explained above, Gaia's mode of operation \footnote{A magnitude--limited, 
or better, $S/N$ threshold–-limited survey, uneven
coverage, including time sampling and scanning geometry, depending
on ecliptic latitude.} is such that there cannot be any
optimization to the case of extrasolar planets. The fundamental
requirement, i.e. to have sufficient astrometric accuracy at
magnitudes brighter than $V =13$, was established at time of
the science case definition.\\
Since little can be done with the photometric and spectroscopic
capabilities aboard the satellite, which cannot compete with
present and planned ground-based facilities for very 
high precision radial-velocity measurements (Pepe \& Lovis 2008) 
and space-borne observatories devoted to ultra-high precision transit photometry 
(e.g., Sozzetti et al. 2010), the potential contribution of
Gaia to exoplanets science must be purely gauged in terms of its
astrometric capabilities. 

\subsection{The Gaia Double-Blind Tests Campaign}

A number of authors have tackled the problem of
evaluating the sensitivity of the astrometric technique required to
detect extrasolar planets and reliably measure their orbital elements and
masses (Sozzetti 2005, and references therein). Those works mostly relied on simplifying assumptions
with regard to {\em a} the error models to be applied to the data (e.g.,
simple Gaussian distributions, perfect knowledge of the instruments) and
{\em b} the analysis procedures to be adopted for orbit reconstruction (mostly ignoring the problem
of identifying adequate configurations of starting values from scratch).
The two most recent exercises on this subject (Casertano et al. 2008; Traub et al. 2009)
have revisited earlier findings using a more realistic double-blind protocol.
In this particular case, several teams of ``solvers'' handled simulated datasets of
stars with and without planets and independently defined detection tests,
with levels of statistical significance of their choice, and orbital fitting algorithms,
using any local, global, or hybrid solution method that they judged was best. The solvers
were provided no information on the actual presence of planets around a given target.

\begin{figure}
\centering
\includegraphics[width=0.9\textwidth]{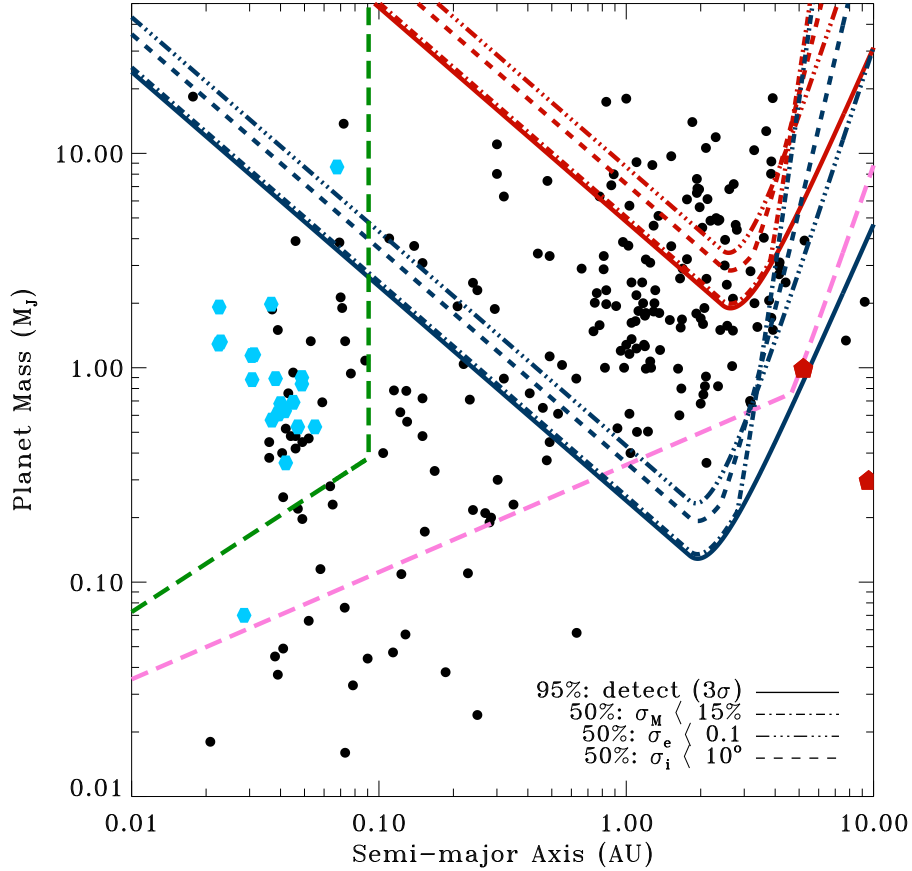}
\caption{Gaia discovery space for planets of given mass and orbital radius compared
to the present-day sensitivity of other indirect detection methods, namely Doppler
spectroscopy and transit photometry. Red curves of different styles
(for completeness in planet detection and orbit measurement to given accuracy)
assume a 1-$M_\odot$ G dwarf primary at 200 pc,
while the blue curves are for a 0.5-$M_\odot$ M dwarf at 25 pc. The radial velocity
curve (pink line) is for detection at the $3\times\sigma_\mathrm{RV}$ level, assuming
$\sigma_\mathrm{RV} = 3$ m s$^{-1}$, $M_\star = 1 M_\odot$, and 10-yr survey duration.
For transit photometry (green curve),
$\sigma_V = 5$ milli-mag, $S/N = 9$, $M_\star=1$ $M_\odot$, $R_\star = 1$ $R_\odot$,
uniform and dense ($> 1000$ datapoints) sampling.
Black dots indicate the inventory of exoplanets as of October 2007. Transiting systems
are shown as light-blue filled pentagons. Jupiter and Saturn are also shown as red pentagons. 
{\it Credits: Casertano et al. 2008}}
\label{Gaiafig}
\end{figure}

In the large-scale, double-blind test (DBT) campaign carried out to estimate the potential of
Gaia for detecting and measuring planetary systems, Casertano et al. (2008) showed that
{\em a}
planets with $\alpha\simeq 6\sigma$ (where $\sigma$ is the single-measurement error)
and orbital periods shorter than the nominal 5~yr mission
lifetime could be accurately modeled, and {\em b} for favorable configurations of two-planet systems
with well-separated periods (both planets with $P\leq 4$ yr and $\alpha/\sigma\geq 10$,
redundancy over a factor of 2 in the number of observations) it would be possible to carry out
meaningful coplanarity tests, with typical uncertainties on the mutual inclination angle of $\leq 10$ deg.
Both subtle differences as well as significant discrepancies were found in the orbital solutions carried
out by different solvers. This constitutes further evidence that the convergence of non-linear
fitting procedures and the quality of orbital solutions (particularly for multiple systems and for
systems with small astrometric signals) can be significantly affected by the choice of the
starting guesses for the parameters in the orbital fits, by the adoption of different statistical
indicators of the quality of a solution, and by varied levels of significance of the latter.

\begin{figure}
\centering
\includegraphics[width=.31\textwidth,angle=270.]{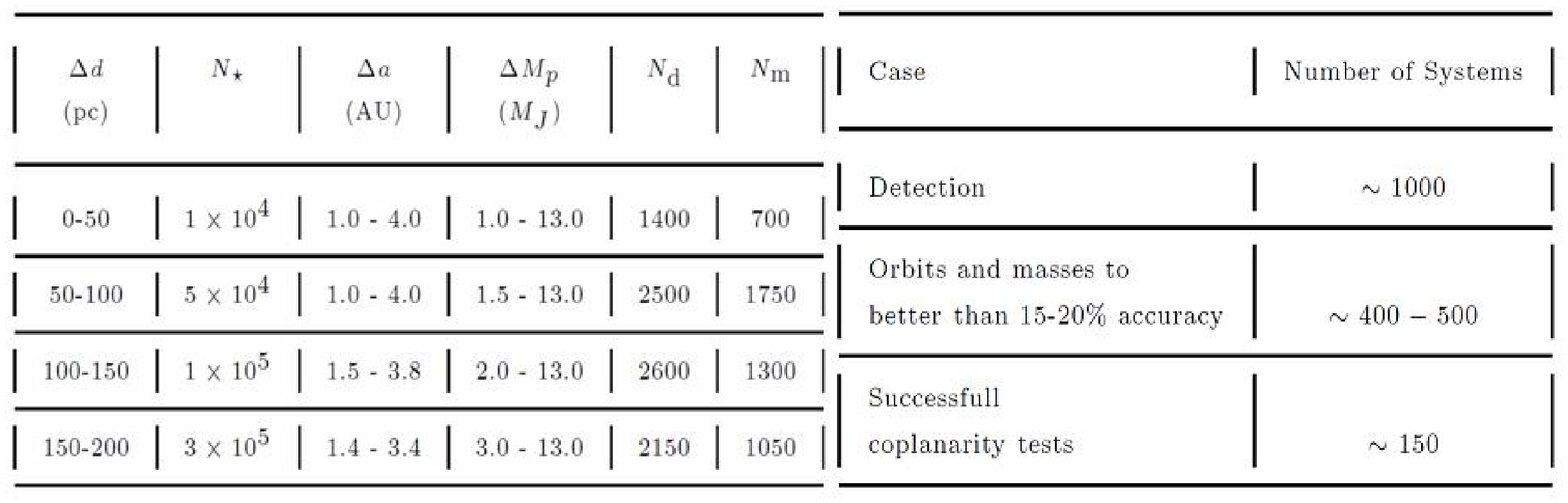}
\caption{Left: Number of giant planets that could be detected and measured by Gaia, as a
function of increasing distance. Starcounts are obtained using models of
stellar population synthesis (Bienaym\'e et al. 1987), while the Tabachnik \& Tremaine (2002) model
for estimating planet frequency as a function of mass and orbital period is used.
Right: Number of planetary systems that Gaia could potentially detect, measure,
and for which coplanarity tests could be carried out successfully. {\it Credits: Casertano et al. 2008}.}
\label{Gaiatab}
\end{figure}

Overall, the authors concluded that Gaia could discover and measure massive giant planets ($M_{\mathrm{p}} \geq $2--3 $M_\mathrm{J}$)
with $1<a<4$ AU orbiting solar-type stars as far as the nearest star-forming regions,
as well as explore the domain of Saturn-mass planets with similar
orbital semi-major axes around late-type stars within 30--40~pc (see Figure~\ref{Gaiafig}).
These results can be
used to infer the number of planets of given mass and orbital separation that can be detected and measured by Gaia,
using Galaxy models and the current knowledge of exoplanet frequencies. By inspection of the tables in
Figure~\ref{Gaiatab}, one then finds that Gaia's main strength will be its ability to accurately measure
orbits and masses for thousands of giant planets, and to perform coplanarity measurements for a few
hundred multiple systems with favorable configurations.

\section{The Gaia Legacy}

Gaia's main contribution to exoplanet science will be its unbiased census of planetary systems 
orbiting hundreds of thousands nearby ($d< 200$ pc), relatively bright ($V \leq 13$) stars 
across all spectral types, screened with constant astrometric sensitivity. 
The Gaia data have the potential to: 

\begin{itemize}
\item[a)] significantly refine our understanding of the statistical properties
of extrasolar planets: the predicted database of several
thousand extrasolar planets with well-measured properties will allow
for example to test the fine structure of giant planet parameters
distributions and frequencies, and to investigate their possible changes
as a function of stellar mass, metallicity, and age with unprecedented resolution;

\item[b)] help crucially test theoretical models of
gas giant planet formation and migration: for example, specific predictions
on formation time-scales and the role of varying metal content in the
protoplanetary disk will be probed with unprecedented statistics thanks
to the thousands of metal-poor stars and hundreds of young stars
screened for giant planets out to a few AUs ; 

\item[c)] achieve key improvements in our comprehension of important aspects of
the formation and dynamical evolution of multiple-planet systems: 
for example, the measurement of orbital parameters
for hundreds of multiple-planet systems, including meaningful coplanarity
tests will allow to discriminate between various proposed mechanisms
for dynamical interaction; 

\item[d)] aid in the understanding of
direct detections of giant extrasolar planets: for example, actual mass estimates and full orbital
geometry determination for suitable systems will
inform direct imaging surveys about the epoch and location of maximum 
brightness, in order to
estimate optimal visibility, and will help in the modeling and interpretation
of giant planets' phase functions and light curves;

\item[e)] provide important supplementary data for the optimization of the
target selection for future observatories aiming at the direct detection 
and spectral characterization of 
habitable terrestrial planets: for example, all F-G-K-M stars within the useful volume ($\sim 25$ pc)
will be screened for Jupiter- and Saturn-sized planets out to several AUs, and
these data will help probing the long-term dynamical stability of their
Habitable Zones, where terrestrial planets may have formed, and maybe found.
\end{itemize}

\section{Conclusions}

The largest compilation of astrometric orbits of giant planets
(in many cases signposts of more interesting systems!), unbiased across all spectral types up to $d\simeq200$ pc, 
will allow Gaia to crucially contribute to several aspects of planetary systems astrophysics 
(formation theories, dynamical evolution), in combination with present-day and future 
extrasolar planet search programs.

\acknowledgements

Financial support from the Italian Space Agency (ASI) through 
contract I/037/08/0 (Gaia Mission - The Italian Participation in
DPAC) is gratefully acknowledged.


\end{document}